\documentclass[12pt]{iopart}
\usepackage{graphicx}

\providecommand{\gtrsim}{\:\raisebox{.25ex}{$>$}\hspace*{-.75em}
\raisebox{-.93ex}{$\sim$}\:}
\providecommand{\lesssim}{\:\raisebox{.25ex}{$<$}\hspace*{-.75em}
\raisebox{-.93ex}{$\sim$}\:}

\begin{document}

\title[Radiation-dominated implosion of tungsten plasma]{1D study of radiation-dominated implosion of a cylindrical tungsten plasma column}

\author{M~M~Basko$^{1,2}$, P~V~Sasorov$^1$, M~Murakami$^2$, V~G~Novikov$^3$ and A~S~Grushin$^3$}

\address{$^1$ Alikhanov Institute for Theoretical and Experimental
Physics, Moscow, Russia}
\address{$^2$ Institute of Laser Engineering, Osaka University, Suita, Osaka 565-0871, Japan}
\address{$^3$ Keldysh Institute of Applied Mathematics, Moscow, Russia}

\ead{basko@itep.ru}


\begin{abstract}
Spectral properties of the x-ray pulses, generated by perfectly uniform cylindrical implosions of tungsten plasma with parameters typical of wire array z-pinches, are investigated under the simplifying assumption that the final stage of the kinetic-to-radiant energy conversion is not affected by the magnetic field. The x-ray emission is shown to be generated within a narrow (sub-micron) radiation-dominated stagnation shock front with a ``supercritical'' amplitude. The structure of the stagnation shock is investigated by using two independent radiation-hydrodynamics codes, and by constructing an approximate analytical model. The x-ray spectra are calculated for two values of the plasma column mass, 0.3~mg~cm${}^{-1}$ and 6~mg~cm${}^{-1}$, with a newly developed two-dimensional radiation-hydrodynamics code RALEF-2D. The hard component of the spectrum (with a blackbody-fit temperature of 0.5--0.6~keV for the 6-mg~cm${}^{-1}$ mass) originates from a narrow peak of the electron temperature inside the stagnation shock. The softer main component emerges from an extended halo, where the primary shock radiation is reemitted by colder layers of the imploding plasma.  Our calculated x-ray spectrum  for the 6-mg~cm${}^{-1}$ tungsten column agrees well with the published Sandia experimental data (Foord \textit{et al} 2004 \textit{Phys.\ Rev.\ Lett.} \textbf{93}, 055002).
\end{abstract}


\submitto{\PPCF}

\maketitle

\section{Introduction}

Wire array z-pinches proved to be one of the most efficient and practical way to generate multi-terawatt pulses of quasi-thermal x-rays with duration of a few nanoseconds \cite{SpDe.98,DeDo.98}. This implies reach potential for many applications, in particular, as an attractive driver for inertial confinement fusion (ICF) \cite{MaSw.05,CuVe.06}. From theoretical point of view, the wire array z-pinch is a complex physical phenomenon: its adequate modelling requires sophisticated multi-dimensional magnetohydrodynamic (MHD) simulations of a complex plasma-metal configuration, whose dynamics at a later stage is strongly influenced by radiative processes \cite{ChLe.01,ChLe.04,JeCu.10}. 

The focus of this paper is on a single specific aspect of this phenomenon, namely, on the physics of conversion of the kinetic energy of the imploding pinch into the emerging pulse of x-rays. We solve an idealized one-dimensional (1D) problem where a high-velocity cylindrical implosion of tungsten plasma is stopped according to the laws of pure radiation hydrodynamics, with a possible role of  MHD effects neglected. We believe that, despite a seemingly oversimplified statement of the problem, thus obtained results 
will prove to be useful for interpretation and subsequent modelling of the x-ray pulses and their spectra in wire array z-pinches optimized for maximum x-ray power. The emission that we calculate should primarily refer to the main part of the observed pulses, i.e.\  to the x-ray flux in a narrow time window around the peak of the x-ray power with the full width at half maximum (FWHM) roughly equal to its rise time ($\simeq 5$~ns in a typical 18--19~MA shot on the Z machine at Sandia \cite{StIv.04}). Note that thus defined main pulse may only contain about 50\% of the total x-ray emitted energy \cite{StIv.04}.

Our analysis is based on the assumption that practically all the energy radiated in the main pulse originates from the kinetic energy of the imploding plasma. Such a premise is corroborated by the latest 3D MHD simulations of imploding wire arrays \cite{JeCu.10}. To simplify the treatment, we do not consider the acceleration stage of the plasma implosion and start with an initial state at maximum implosion velocity. Hence, we do not have to consider the $\bf{j} \times \bf{B}$ force which drives the implosion (i.e.\ accelerates the plasma but generates little entropy) because the entire kinetic energy of the implosion can be simply prescribed at the initial state. The resistive (Ohmic) dissipation of the electromagnetic energy was found to be negligible anyway \cite{JeCu.10}. More difficult is to justify our neglect of the MHD effects at the final stage of plasma deceleration and x-ray generation. Although some additional arguments to this point are given in the summary, we must admit that full clarification of this issue remains for future work.

Under the assumptions made, we find that the kinetic energy of the implosion is converted into radiation when the plasma passes through a stagnation shock near the axis. In wire arrays with powerful x-ray emission the stagnation shock falls into the category of ``supercritical'' radiation-dominated (RD) shock fronts \cite{ZeRa67}. Such fronts are characterized by an extremely small width, and their thermal structure is primarily controlled by emission and transport of thermal radiation. Our results imply that adequate modelling of the temperature and density profiles across the narrow stagnation shock front is key to understanding the x-ray spectra emitted by powerful wire array z-pinches.

The paper is organized as follows. In section~\ref{s:pr} we describe the initial state of the simulated plasma flow; section~\ref{s:code} elaborates on the two numerical codes used for simulations. Section~\ref{s:shock} is devoted to the detailed analysis of the stagnation shock structure: we construct an approximate analytical model, which is corroborated by numerical simulations and allows simple evaluation of the key plasma parameters in the shock front. In section~\ref{s:ph} we present the calculated spectra of the x-ray pulses, radial profiles of the spectral optical depth, spectral x-ray images of the plasma column. The emergent x-ray spectra have been calculated for two values of the linear pinch mass, 0.3~mg~cm${}^{-1}$ and 6~mg~cm${}^{-1}$, by employing a newly developed two-dimensional (2D) radiation-hydrodynamics code RALEF-2D.

\section{\label{s:pr} Initial configuration}

We choose the simplest initial configuration that allows reproduction of the principal characteristics of the observed x-ray pulses. We start with  a cylindrical shell of tungsten plasma, converging onto the pinch axis $r=0$ with an initial implosion velocity $u(0,r)=-U_0$ ($U_0>0$) constant over the shell mass.  The imploding shell is supposed to have sharp boundaries at $r=r_1(t)$ and $r= r_1(t)+ \Delta_0$. Once the radial velocity peaks at $U_0$, the implosion can be treated as ``cold'' in the sense that the plasma internal energy is small compared to its kinetic energy, the role of pressure forces is negligible, and the shell thickness freezes at a constant value $\Delta_0$. We begin our simulations at time $t=0$ when the inner shell edge arrives upon the axis, i.e.\ when $r_1=r_1(0) =0$.

The density distribution across the imploding shell is assumed to have been uniform at earlier times, when the inner shell radius was $r_1(t) \gg \Delta_0$. In a cold implosion this leads to the initial radial density profile of the form
\begin{equation}\label{rho_0=}
  \rho_0(r) = \left(\frac{m_0}{2\pi \Delta_0}\right) \frac{1}{r},
\end{equation}
where $m_0$ is the linear (per unit cylinder length) mass of the shell.

In this paper we present simulations for two cases, namely, case~A (referring to the 5-MA Angara-5-1 machine in Troitsk, Russia) and case~Z (referring to the 20-MA Z accelerator at Sandia, USA). In both cases we used the same values of the implosion velocity and shell thickness,
\begin{equation}\label{U_Del=}
  U_0 = 400 \mbox{ km~s${}^{-1}$} =4\times 10^7 
  \mbox{ cm~s${}^{-1}$},  \qquad
  \Delta_0 =2\mbox{ mm}.
\end{equation}
The peak implosion velocity of 400~km~s${}^{-1}$ has been inferred from the experimental data for optimized shots on both the Angara-5-1 \cite{AlGr.04} and the Z machines \cite{CuWa.05}, and confirmed by numerical simulations \cite{ChLe.01,JeCu.10}. For the given $U_0$, the shell thickness $\Delta_0$ is set equal to 2~mm to conform with the observed x-ray pulse duration of 5~ns (FWHM) \cite{SpDe.98,DeDo.98,GrZu.04,GrZu.06}. Note that if, in addition, we assumed a 100\% instantaneous conversion of the kinetic energy into x-rays, we would obtain a rectangular x-ray pulse of duration
\begin{equation}\label{t_0=}
  t_0 = \frac{\Delta_0}{U_0} =5 \mbox{ ns}
\end{equation}
with the top nominal power
\begin{equation}\label{P_0=}
  P_0 = \frac{m_0 U_0^3}{2\Delta_0}
\end{equation}
(per unit cylinder length).

Thus, the only parameter that differs between the cases A and Z is the linear mass $m_0$ of the imploding shell. In simulations we used the values
\begin{equation}\label{m_0=}
  m_0  = \left\{ \begin{array}{ll}
   0.3 \mbox{ mg~cm${}^{-1}$}, & \mbox{case A}, \\
  6.0  \mbox{ mg~cm${}^{-1}$}, & \mbox{case Z},
  \end{array} \right.
\end{equation}
which are representative of a series of optimized (with respect to the peak power and total energy of the x-ray pulse) experiments at a 3~MA current level on Angara-5-1 \cite{GrZu.06}, and at a 19~MA current level on Z \cite{StIv.04}. These two values of $m_0$ correspond to the nominal powers
\begin{equation}\label{P_0==}
  P_0 = \left\{ \begin{array}{ll}
   4.8 \mbox{ TW~cm${}^{-1}$}, & \mbox{case A}, \\
  96  \mbox{ TW~cm${}^{-1}$}, & \mbox{case Z},
  \end{array} \right.
\end{equation}
which are close to the peak x-ray powers measured in the corresponding experiments.

The final parameter needed to fully specify the initial state is the initial temperature $T_0$ of the imploding plasma. In both cases we used the same value $T_0 =20$~eV, which falls in the 10--30~eV range  inferred from the theory of plasma ablation in multi-wire arrays \cite{AlBr.01,YuOl.07,SaOl.08} and confirmed by direct MHD simulations of the wire-corona plasma \cite{ChLe.01}.  The sound velocity in a 20-eV tungsten plasma, $c_{\mathrm{s}} \approx (\mbox{0.5--1.0})\times 10^6$~cm~s${}^{-1}$, implies implosion Mach numbers as high as $U_0/c_{\mathrm{s}} \approx 40$--80 --- which fully justifies the above assumption of a cold implosion.

\section{\label{s:code} The  DEIRA and the RALEF-2D
codes}

Numerical simulations have been performed with two numerical codes that are based on different numerical techniques and include fully independent models of all physical processes, namely, with a 1D three-temperature (3T) code \mbox{DEIRA} \cite{Bas90,DEIRA}, and a 2D radiation-hydrodynamics code RALEF-2D \cite{RALEF}. Because of strongly differing physical models and numerical capabilities, the results obtained with these two codes are to a large extent complimentary to one another. The 2D \mbox{RALEF} code was used to simulate our 1D problem simply because we had no adequate 1D code with spectral radiation transport at hand.

\subsection{The DEIRA code}

The 1D 3T \mbox{DEIRA} code was originally written  to simulate ICF targets \cite{Bas90}. It is based on one-fluid Lagrangian hydrodynamics with a tensor version of the Richtmyer artificial viscosity and different electron, $T_{\mathrm{e}}$, and ion, $T_{\mathrm{i}}$, temperatures. Included also are the electron and the ion thermal conduction, as well as the ion physical viscosity. The model for the electron and the ion conduction coefficients is based on the Spitzer formula, modified in such a way as to match the experimental data near normal conditions \cite{DEIRA}. The equation of state is based on the average ion model \cite{Bas85}, which accounts for both the thermal and the pressure ionization at high temperatures and/or densities, as well as for realistic properties of materials near normal conditions.

Energy transport by thermal radiation is described by a separate diffusion equation for the radiative energy density $\rho \epsilon_{\mathrm{r}} = a_{\mathrm{S}} T_{\mathrm{r}}^4$, expressed in terms of a separate radiation temperature $T_{\mathrm{r}}$; here $a_{\mathrm{S}}$ is the Stefan constant. The energy relaxation between the electrons and the radiation is expressed in terms of the Planckian mean absorption coefficient $k_{\mathrm{Pl}}$, while the radiation diffusion coefficient is coupled to the Rosseland mean $k_{\mathrm{R}}$. The absorption coefficients $k_{\mathrm{Pl}}$ and  $k_{\mathrm{R}}$ are evaluated in-line as a combined contribution from the free-free, bound-free and bound-bound electron transitions plus the Thomson scattering. For the bound-bound and bound-free transitions, an approximate model, based on the sum rule for the dipole oscillator strengths \cite{ImMi86}, is used. The subset of the three energy equations is solved in a fully implicit manner by linearizing with respect to the three unknown temperatures $T_{\mathrm{e}}$, $T_{\mathrm{i}}$, and $T_{\mathrm{r}}$.

\subsection{The RALEF-2D code}

RALEF-2D (Radiative Arbitrary Lagrangian-Eulerian Fluid dynamics in two Dimensions) is a new radiation-hydrodynamics code, whose development is still underway \cite{RALEF}. Its hydrodynamics module is based on the upgraded version of the \mbox{CAVEAT} hydrodynamics package \cite{CAVEAT}. The one-fluid one-temperature hydrodynamic equations are solved in two spatial dimensions [in either Cartesian $(x,y)$ or axisymmetric $(r,z)$ coordinates] on a multi-block structured quadrilateral grid by a second-order Godunov-type numerical method. An important ingredient is the rezoning-remapping algorithm within the Arbitrary Lagrangian-Eulerian (ALE) approach to numerical hydrodynamics. The original mesh rezoning scheme, based on the Winslow equipotential method \cite{Win81}, proved to be quite efficient for the interior of the computational domain, if the mesh is smooth along the boundaries; in RALEF, a new high-order method for rezoning block boundaries has been implemented to this end.

New numerical algorithms for thermal conduction and radiation transport have been developed within the unified symmetric semi-implicit approach \cite{LiGl85} with respect to time discretization. The algorithm for thermal conduction is a conservative, second-order accurate symmetric scheme on a 9-point stencil \cite{BaMa.09}. Radiation energy transport is described by the quasi-static transfer equation
\begin{equation}\label{eq:r-transp=}
  \mathbf{\Omega} \cdot \nabla I_{\nu} = k_{\nu} \left(B_{\nu}
  -I_{\nu}\right)
\end{equation}
for the spectral radiation intensity $I_{\nu} =I_{\nu}(t,\mathbf{x},\mathbf{\Omega})$; the term $c^{-1} \partial I_{\nu}/\partial t$, where $c$ is the speed of light, is neglected. A non-trivial issue for spatial discretization of equation~(\ref{eq:r-transp=}) together with the radiative heating term
\begin{equation}\label{eq:Q_r=}
  Q_{\mathrm{r}} = -\mathop{\mathrm{div}} \int\limits_0^{\infty} d\nu
 \int\limits_{4\pi}I_{\nu}\, \mathbf{\Omega}\;  d\mathbf{\Omega}
\end{equation}
in the hydrodynamic energy equation, is correct reproduction of the diffusion limit on distorted non-orthogonal grids \cite{LaMo.87}. In our scheme, we use the classical $S_n$ method to treat  the angular dependence of the radiation intensity $I_{\nu}(t,\mathbf{x},\mathbf{\Omega})$, and the method of short characteristics \cite{DeVo02} to integrate equation~(\ref{eq:r-transp=}). The latter has a decisive advantage that every grid cell automatically receives the same number of light rays.  Correct transition to the diffusion limit is achieved by special combination of the first- and second-order interpolation schemes in the finite-difference approximations to equations~(\ref{eq:r-transp=}) and (\ref{eq:Q_r=}). More details on the numerical scheme for radiation transfer are to be published elsewhere.

In the present work we used the equation of state, thermal conductivity and spectral opacities provided by the \mbox{THERMOS} code \cite{THERMOS}, which has been developed at the Keldysh Institute of Applied Mathematics (Moscow). The spectral opacities are generated by solving the Hatree-Fock-Slater equations for plasma ions under the assumption of equilibrium level population. In combination with the equilibrium Planckian intensity $B_{\nu}$, used in (\ref{eq:r-transp=}) as the source function, the latter means that we treat radiation transport in the approximation of local thermodynamic equilibrium (LTE) --- which is justified for relatively dense and optically thick plasmas considered here.

\begin{figure}[hbt!]
\includegraphics*[width=80mm]{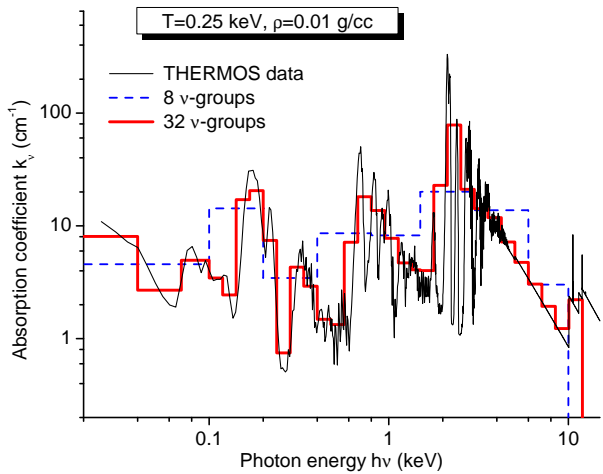}
\caption{\label{f:knu} (Colour online) Spectral absorption coefficient $k_{\nu}$ [cm${}^{-1}$] of tungsten at $\rho =0.01$~g~cm${}^{-3}$, $T=250$~eV used in the present simulations: shown are the original data from the \mbox{THERMOS} code (thin solid curve) together with the group-averaged values for 8 (dashed) and 32 (thick solid) selected spectral groups.}
\end{figure}

The transfer equation (\ref{eq:r-transp=}) is solved numerically for a selected number of discrete spectral groups $[\nu_j, \nu_{j+1}]$, with the original \mbox{THERMOS} absorption coefficients $k_{\nu}$ averaged inside each group $j$ by using the Planckian weight function. Two different sets of frequency groups are prepared for each code run: the primary set with a smaller number of groups (either 8 or 32 in the present simulations) is used at every time step in a joint loop with the hydrodynamic module, while the secondary (diagnostics) set with a larger number of groups (200 in the present simulations) is used in the post-processor regime at selected times to generate the desired spectral output data. An example of the spectral dependence of $k_{\nu}$, provided by the \mbox{THERMOS} code for a tungsten plasma at $\rho =0.01$~g~cm${}^{-3}$, $T=250$~eV, is shown in figure~\ref{f:knu} together with the corresponding group averages used in the RALEF simulations.

\subsection{Numerical setup for the \mbox{RALEF} simulations}

To test the sensitivity of the results with respect to spectral radiation transport, we did our simulations with two selections for the primary set of frequency groups, namely,
\begin{itemize}
\item  with 8  groups delimited by the photon energies
\begin{equation}\label{nu_j(8g)=}
  h\nu_j = 10^{-3}, \;  0.1, \;  0.2,\;  0.4,\;  0.8,\;  1.5,\;  3.0,\;
  6.0,\;  10.0 \;
  \mbox{ keV},
\end{equation}

\item  and with 32  groups delimited by
\begin{eqnarray}\label{nu_j(32g)=}
  h\nu_j &=& 10^{-3}, \;  0.02, \;  0.04,\;  0.07,\;  0.1,\;  0.119,\;  0.1414,\;
   \nonumber \\ &&
   0.168,\;  0.20, \; 0.238, \; 0.2828, \; 0.336, \; 0.4, \; 0.476,
   \nonumber \\ &&
   0.5656, \; 0.672, \; 0.8, \; 0.952, \; 1.1312, \; 1.344, \; 1.5,
   \nonumber \\ &&
   1.785, \; 2.121, \; 2.52, \; 3.0, \; 3.57, \; 4.242, \; 5.04,\;
   \nonumber \\ &&
   6.0, \; 7.14, \; 8.484, \; 10.08, \; 12.0 \;
  \mbox{ keV}.
\end{eqnarray}

\end{itemize}
The 200 spectral groups of the secondary (diagnostics) frequency set were equally spaced along $\ln(h\nu)$ between $h\nu_1=0.01$~keV and $h\nu_{201} =10$~keV. The angular dependence of the radiation intensity was calculated with the $S_{14}$ method, which offers 28 discrete ray directions per octant.

The simulated region occupied one quadrant $0\leq \phi \leq 90^{\circ}$ of the azimuth angle $\phi$ with reflective boundaries along the $x$- and $y$-axes. Near the geometrical centre $x=y=0$, a rigid transparent wall was placed at $r=r_0 =10$~$\mu$m with the boundary conditions of $u(t,r_0)=0$ and zero thermal flux. Thermal radiation passed freely through this cylindrical wall and was reflected by the two perpendicular reflective boundaries. Two variants of the initial polar mesh were used: a $n_{\phi} \times n_{\mathrm{r}} = 50\times 250$ mesh in case~A, and a $n_{\phi} \times n_{\mathrm{r}} = 60\times 600$ mesh in case~Z.  At the outer boundary (initially at $r=R_0 =2$~mm), the boundary conditions of zero external pressure and zero incident radiation flux were applied.

\section{\label{s:shock} Stagnation shock}

\subsection{General picture}

\begin{figure}[htb!]
\includegraphics*[height=60mm]{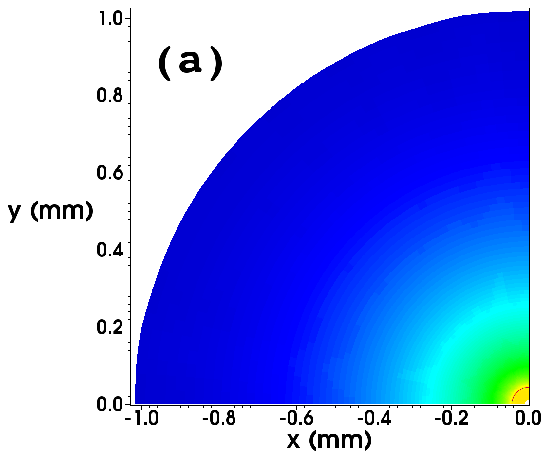} 
\includegraphics*[height=60mm]{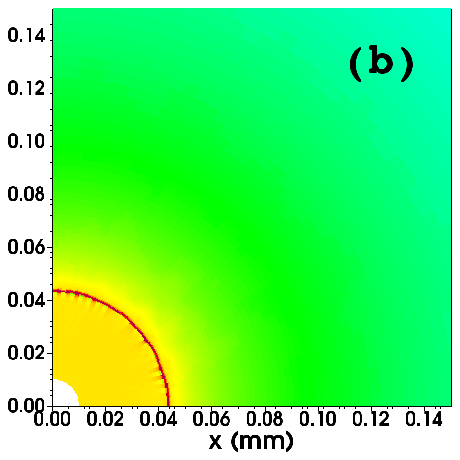}
\caption{\label{f:2Dview} (Colour) 2D contour maps of matter temperature  at $t=3$~ns in case~A (as calculated with the \mbox{RALEF} code): frame~(b) is a blow up of the central part of the full view~(a). Colour represents matter temperature $T$. A thin dark-red circle at $r= 43$~${}\mu$m marks the position of the stagnation shock.}
\end{figure}

Upon arrival at the axis, the imploding plasma comes to a halt passing through a stagnation shock. In our situation the specific nature of this shock is defined by the dominant role of the radiant energy exchange. A detailed general analysis of the structure of such RD shock waves is given in \cite[Ch.~VII]{ZeRa67}. Perhaps the most salient feature of an RD shock with a supercritical amplitude is a very narrow local peak of matter temperature immediately behind the density jump \cite{ZeRa67}. This temperature peak manifests itself as a hair-thin bright circle at $r=r_{\mathrm{s}} \approx 43$~${}\mu$m on the 2D temperature plot in figure~\ref{f:2Dview}.

\begin{figure}[htb!]
\includegraphics*[width=80mm]{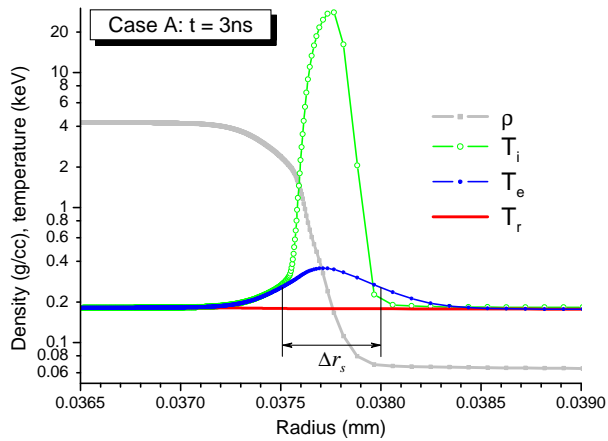}
\caption{\label{f:deiA} (Colour online) Density and temperature profiles across the stagnation shock in case~A as calculated with the \mbox{DEIRA} code for $t=3$~ns.  The effective width $\Delta r_{\mathrm{s}}$ of the shock front is defined as the FWHM of the local peak of the electron temperature $T_{\mathrm{e}}$. The velocity profile in the shock frame can be easily restored from the density plot and the condition $\rho v = constant$, which is quite accurately observed in the displayed region.}
\end{figure}

\begin{figure}[htb!]
\includegraphics*[width=80mm]{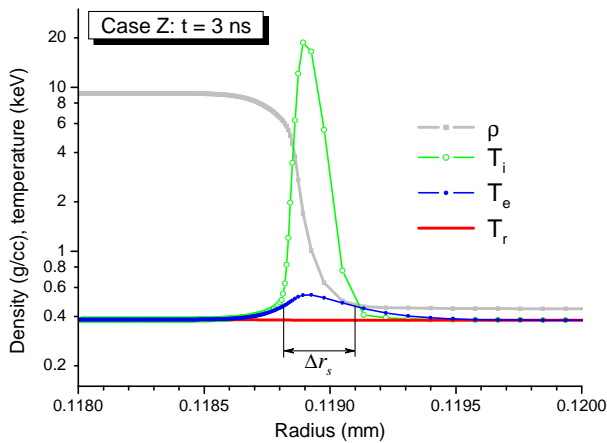}
\caption{\label{f:deiZ} (Colour online) Same as figure~\ref{f:deiA} but for case~Z.}
\end{figure}

To achieve an adequate numerical resolution of the RD shock front, one needs a very fine grid that can only be afforded in 1D simulations. Figures~\ref{f:deiA} and \ref{f:deiZ} show the density and temperature profiles across the stagnation shock at $t=3$~ns as calculated with the 1D \mbox{DEIRA} code on a uniform Lagrangian mesh with 20~000 mass intervals. If we define the shock-front width $\Delta r_{\mathrm{s}}$ to be the FWHM of the hump on the $T_{\mathrm{e}}$ profile, we obtain $\Delta r_{\mathrm{s}} =0.5$~${}\mu$m in case~A, and $\Delta r_{\mathrm{s}} =0.3$~${}\mu$m in case~Z. The peak values of the electron temperature are calculated to be $T_{\mathrm{ep}}=0.35$~keV in case~A, and $T_{\mathrm{ep}}=0.54$~keV in case~Z.

In a model where one distinguishes between the electron and the ion temperatures but ignores viscosity and ion heat conduction, the shock front has a discontinuity in the density and $T_{\mathrm{i}}$ profiles. In figures~\ref{f:deiA} and \ref{f:deiZ} this discontinuity is smeared out (roughly over 3 mesh cells) by the artificial Richtmyer-type viscosity, present in the \mbox{DEIRA} code. The electron temperature $T_{\mathrm{e}}$, which exhibits a prominent hump over the virtually constant radiation temperature $T_{\mathrm{r}}$, is continuous because the electron thermal conduction plays a significant role. Clearly, the plasma inside the dense part of the $T_{\mathrm{e}}$ hump must be intensely loosing energy via thermal radiation.

Figures~\ref{f:deiA} and \ref{f:deiZ} demonstrate prominent peaks of the ion temperature $T_{\mathrm{i}}$, whose maximum values $T_{\mathrm{ip}} \approx 20$~keV significantly exceed the peak electron temperature $T_{\mathrm{ep}}$. This fact, however, turns out to be rather insignificant for the radiative properties of the imploding plasma. Indeed, if we assume that the kinetic energy of the infalling plasma is fully converted into the ion thermal energy within a density-temperature discontinuity and ignore the plasma preheating before the shock, we calculate an after-shock ion temperature of
\begin{equation}\label{T_i+=}
  T_{\mathrm{i}+} =\frac{\gamma-1}{2} m_{\mathrm{i}} U_0^2 =\frac{1}{3} m_{\mathrm{i}} U_0^2 =   \mbox{100 keV};
\end{equation}
here $\gamma =5/3$ is the adiabatic index of the ideal gas of plasma ions, $m_{\mathrm{i}}$ is the mass of a tungsten ion. The \mbox{DEIRA} simulations demonstrate much lower peak ion temperatures because the preheating of the pre-shock plasma electrons, followed by their adiabatic compression in the density jump, consumes a large portion of the initial ion kinetic energy (in a collisionless manner via ambipolar electric fields). As a consequence, even before the collisional electron-ion relaxation sets in, the post-shock electrons with a temperature of $T_{\mathrm{e}} \approx 0.4$~keV already contain almost twice as much energy as the post-shock ions with a temperature of $T_{\mathrm{i}} \approx 20$~keV. Hence, the subsequent collisional electron-ion relaxation does not significantly affect the $T_{\mathrm{e}}$ profile. This fact has also been verified directly: having performed additional \mbox{DEIRA} runs in the 2T mode (i.e.\ assuming $T_{\mathrm{e}}=T_{\mathrm{i}}=T$), we obtained $T$ profiles that were hardly distinguishable from the $T_{\mathrm{e}}$ profiles in figures~\ref{f:deiA} and \ref{f:deiZ} (the difference between the peak values $T_{\mathrm{p}}$ and $T_{\mathrm{ep}}$ did not exceed 3\%). Thus, the approximation of a single matter temperature $T=T_{\mathrm{e}}=T_{\mathrm{i}}$, used in the 2D \mbox{RALEF} code, is well justified for our problem.

\subsection{Analytical model}

The theory of RD shock fronts, developed by Yu.~P.~Raizer \cite{Rai57} and described in his book with Ya.~B.~Zel'dovich \cite{ZeRa67}, applies to planar shock waves in an infinite medium, which eventually absorbs all the emitted photons. We, in contrast, are dealing with a finite plasma mass, which lets out practically all the radiation flux generated at the shock front. In addition, the electron heat conduction, ignored in Raizer's treatment, plays an important role in formation of the temperature profile across the shock front. Hence, we have to reconsider certain key aspects of the Raizer's theory in order to obtain an adequate model for the stagnation shock in imploding z-pinch plasma.

\subsubsection{General relationships}

To construct an analytical model of the plasma flow, we have to make certain simplifying assumptions. First of all, we assume a single temperature $T$ for ions and electrons and employ the ideal-gas equation of state in the form
\begin{equation}\label{EOS=}
  p=A \rho T, \qquad \epsilon = \frac{A}{\gamma-1}\, T,
\end{equation}
where $p$ is the pressure, $\epsilon$ is the mass-specific internal energy, and $A$ and $\gamma> 1$ are constants. The thermodynamic properties of the tungsten plasma in the relevant range of temperatures and densities are reasonably well reproduced with
\begin{eqnarray}\label{A=}
  A &=& \left\{ \begin{array}{ll}
  13 \mbox{ MJ g${}^{-1}$ keV${}^{-1}$} & \mbox{in case A}, \\
  20 \mbox{ MJ g${}^{-1}$ keV${}^{-1}$} & \mbox{in case Z},
  \end{array} \right.  \\ \label{gam=}
  \gamma &=&\left\{ \begin{array}{ll}
  1.29 & \mbox{in case A}, \\ 1.33  & \mbox{in case Z}. \end{array} \right.
\end{eqnarray}

At each time $t$ the entire plasma flow can be divided into three zones: the inner stagnation zone (the compressed core) at $0<r< r_{\mathrm{s}}$ behind the shock front, the shock front itself confined to a narrow layer around $r = r_{\mathrm{s}}$, and the outer layer of the unshocked infalling material at $r> r_{\mathrm{s}}$. In the stagnation zone the plasma velocity is small compared to $U_0$, while the temperature and density are practically uniform and have the final post-shock values of $T=T_1 =T_1(t)$, $\rho= \rho_1 =\rho_1(t)$. Because the plasma flow in the stagnation zone is subsonic, pressure is rapidly equalized by hydrodynamics, while temperature is equalized by efficient radiative heat conduction; note that typical values of the mean Rosseland optical thickness of the stagnant core lie in the range $\tau_{\mathrm{c,Ros}} \simeq 10$--100. Numerical simulations confirm that spatial density and temperature variations across the compressed core do not exceed a few percent.

We identify the shock radius $r_{\mathrm{s}} =r_{\mathrm{s}}(t)$  with the density (and velocity) discontinuity, which is always present in sufficiently strong shocks once a zero physical viscosity is assumed \cite{ZeRa67}. Here and below the term ``shock front'' is applied to a narrow layer with an effective width of $\Delta r_{\mathrm{s}} \ll r_{\mathrm{s}}$, where the matter (electron) temperature $T$ exhibits a noticeable hump above the radiation temperature $T_{\mathrm{r}}$; see figures~\ref{f:deiA} and \ref{f:deiZ}. Outside the shock front at $r>r_{\mathrm{s}}$ lies a broad preheating zone, which extends virtually over the entire unshocked material and has a width well in excess of the shock radius $r_{\mathrm{s}}$. In this region the infalling plasma is preheated due to interaction with the outgoing radiation; in the process, it is also partially decelerated and compressed.

To analyze the structure of the shock front, we use the three basic conservation laws
\begin{eqnarray}\label{bal-m=} 
  \rho v  \equiv  -j = constant, \\ \label{bal-p=} 
  p+ \rho v^2 = constant, \\ \label{bal-e=} 
  \rho v\left(w +\frac{v^2}{2}\right) +S_{\mathrm{e}} +S_{\mathrm{r}} = constant,
\end{eqnarray}
governing a steady-state hydrodynamic flow without viscosity across a planar shock front \cite{ZeRa67}; here
\begin{equation}\label{w=}
  w=\epsilon +\frac{p}{\rho} =\frac{\gamma A}{\gamma-1}\, T
\end{equation}
is the specific enthalpy, $S_{\mathrm{e}}$ and $S_{\mathrm{r}}$ are the energy fluxes due, respectively, to the electron thermal conduction and radiation transport. Equations (\ref{bal-m=})--(\ref{bal-e=}) are written in the reference frame comoving with the shock front: in this frame the plasma velocity $v<0$. To avoid confusion, we use symbol ``$v$'' for the plasma velocity in the shock frame, and symbol ``$u$'' for the plasma velocity in the laboratory frame. The velocity of the shock front in the laboratory frame is  $u_{\mathrm{s}} =\rmd r_{\mathrm{s}}/\rmd t$. Clearly, the planar conservation laws (\ref{bal-m=})--(\ref{bal-e=}) can be applied over a narrow front zone with $|r-r_{\mathrm{s}}| \ll r_{\mathrm{s}}$ but not across the broad preheating zone, where the effects of cylindrical convergence become significant.

\subsubsection{\label{s:score} Parameters of the stagnant core}

\begin{figure}[htb!]
\includegraphics*[width=70mm]{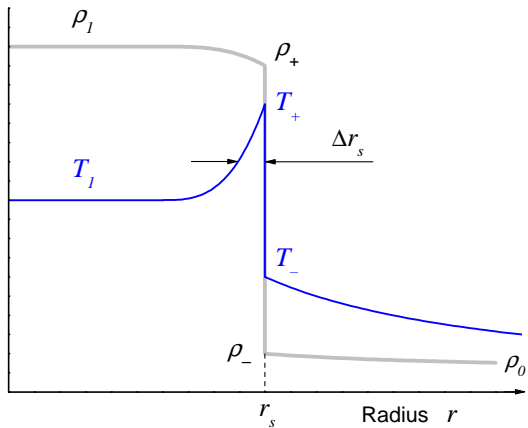}
\caption{\label{f:RDshock} (Colour online) Schematic view of the density, $\rho$, and temperature, $T$, profiles across an RD shock front in the approximation of zero viscosity and heat conduction. The relaxation zones before and after the density jump are due to energy transport by radiation.}
\end{figure}

We begin by deriving a system of equations, from which the parameters of the stagnant plasma core can be evaluated. Although the sought-for quantities formally depend on time $t$, time appears only as a parameter in the final equations. The final post-shock plasma state can be determined in the approximation of zero thermal conduction, which redistributes energy only locally, in the immediate vicinity of the shock front. Without thermal conduction, the density and the temperature should have profiles shown qualitatively in figure~\ref{f:RDshock}: the density jump from $\rho=\rho_-$ to $\rho= \rho_+$ is accompanied by the jump in temperature from $T=T_-$ to $T=T_+$. Immediately behind (in the downstream direction) the temperature peak $T_+$ lies a narrow relaxation zone to the final state $(\rho_1, T_1)$, where the excess thermal energy between the $T_+$ and $T_1$ states is rapidly radiated away.

As was rigorously proven by Ya.~B.~Zel'dovich \cite{Zel57}, the preheating temperature $T_-$ at the entrance into the density jump can never exceed $T_1$. How does $T_-$ compare with $T_1$, depends on whether the RD shock is subcritical or supercritical. A critical amplitude of an RD shock front corresponds to the condition \cite{Rai57}
\begin{equation}\label{crit=}
  \sigma T_1^4 = \rho_1 u_{\mathrm{s}} \epsilon_1 \approx \rho_0U_0
  \frac{AT_1}{\gamma-1}
\end{equation}
that the one-sided radiation energy flux $\sigma T_1^4$ becomes comparable to the hydrodynamic energy flux $\rho_1 u_{\mathrm{s}} \epsilon_1$ behind the shock front; here $\sigma$ is the Stefan-Boltzmann constant. In our case the RD shock becomes supercritical when the post-shock temperature exceeds the critical value of
\begin{equation}\label{T_1cr=}
  T_{1,cr} \approx 0.3\, \rho_{0,\mathrm{g/cc}}^{1/3} \; \mbox{keV},
\end{equation}
where $\rho_{0,\mathrm{g/cc}}$ is the initial density (\ref{rho_0=}) at $r=r_{\mathrm{s}}$ in g~cm${}^{-3}$. A posteriori, having calculated $T_1$ and $r_{\mathrm{s}}$ from the equations given below, we verify that our shock fronts are supercritical. As shown by Yu.~P.~Raizer \cite{Rai57}, when a supercritical RD shock wave propagates in an infinite medium, it has $T_- \approx T_1$. If the optical thickness of the unshocked material is not exceedingly small, this is also true in the case of a finite plasma size; the latter applies to all configurations considered in this paper and is directly confirmed by the profiles in figures~\ref{f:deiA} and \ref{f:deiZ}.

To avoid treatment of the non-planar preheating zone, we make a simplifying assumption  that partial deceleration of the infalling plasma in the preheating zone can be neglected, i.e.\ that one can set $\rho_- =\rho_0$ and $v_- = -(u_{\mathrm{s}}+U_0)$, where $\rho_0$ is calculated from (\ref{rho_0=}) at $r=r_{\mathrm{s}}$, and $v_-$ is the plasma velocity at the entrance into the jump in the shock front frame. As is demonstrated in \S 16 of chapter~VII in \cite{ZeRa67}, for $\gamma-1 \ll 1$ this approximation is accurate to the second order with respect to the small parameter
\begin{equation}\label{eta_1inf=}
  \eta_{1\infty} = (\gamma-1)/(\gamma+1).
\end{equation}
In fact, it has already been used in condition (\ref{crit=}).

Now we can apply equations (\ref{bal-m=}), (\ref{bal-p=}) of mass and momentum balance between the states $\rho_-, T_-$ and $\rho_1, T_1$: \begin{eqnarray}\label{sh-mass=}
  \rho_0(u_{\mathrm{s}}+ U_0) = \rho_1 |v_1| \equiv j, \\ \label{sh-mom=}
 p_- + \rho_0(u_{\mathrm{s}}+U_0)^2 = p_1 +\rho_1 v_1^2;
\end{eqnarray}
here $\rho_0 =\rho_0(r_{\mathrm{s}})$, and $v_1$ is the unknown material velocity behind the  shock front. Because the shock wave propagates over a falling density profile $\rho_0(r) \propto r^{-1}$, a uniform density distribution behind the front implies that the post-shock density $\rho_1(t)$ decreases with time, and material in the stagnation zone expands. The expansion velocity $u$ is small compared with $U_0$ but not with the front velocity $u_{\mathrm{s}}$. As a consequence, we cannot simply put $v_1= -u_{\mathrm{s}}$.

By virtue of (\ref{sh-mass=}) and (\ref{EOS=}),  equation~(\ref{sh-mom=}) can be transformed to
\begin{equation}\label{mom-bal=}
  \eta_1(1-\eta_1)(u_{\mathrm{s}}+U_0)^2 = \frac{p_1}{\rho_1}\left(1-
  \eta_1\frac{T_-}{T_1}\right),
\end{equation}
where
\begin{equation}\label{eta_1-def=}
  \eta_1 \equiv \frac{\rho_0}{\rho_1} =\frac{|v_1|}{u_{\mathrm{s}}+U_0}
\end{equation}
is the inverse of the compression factor. Restricting our treatment to the case of supercritical RD shocks, where $T_- \approx T_1$,  we get \begin{equation}\label{eta_1=}
  \eta_1 = \frac{AT_1}{(u_{\mathrm{s}}+ U_0)^2}.
\end{equation}
In our model $\eta_1$ is a small parameter, which is even smaller than the inverse compression factor $\eta_{1\infty}$ in the infinite-media. Keeping this in mind, in all the algebra below we consistently retain only the zeroth and the first terms with respect to this parameter.

A subtle point here is that we cannot directly use equation (\ref{bal-e=}) of the energy balance across the shock front. Quasi-uniform density and temperature profiles in the stagnation core ensue from the rapid redistribution of thermal energy over the entire mass of this zone by means of radiation. Hence, the post-shock thermal energy calculated from the local  condition (\ref{bal-e=}) may differ considerably from the required average value. To obtain the latter, we use the condition of global energy balance. For a similar reason, we employ the equation of global mass balance to establish the relationship between the radius $r_{\mathrm{s}}$ and the velocity $u_{\mathrm{s}}$ of the shock front.

The total mass $m_{\mathrm{s}}$ of the compressed core can be expressed as 
\begin{equation}\label{m_s=}
  m_{\mathrm{s}} =m_{\mathrm{s}}(t) =\pi r_{\mathrm{s}}^2 \rho_1 = \frac{m_0}{\Delta_0} \left(r_{\mathrm{s}}
  +U_0t \right),
\end{equation}
which, by virtue of (\ref{eta_1-def=}) and (\ref{rho_0=}), yields
\begin{equation}\label{r_s=}
  r_{\mathrm{s}} =\frac{2\eta_1}{1-2\eta_1}\, U_0 t.
\end{equation}
Since $\eta_1$ varies only slowly with time (this can be verified
a posteriori), equation~(\ref{r_s=}) implies
\begin{equation}\label{u_s=}
  u_{\mathrm{s}} \equiv \frac{\rmd r_{\mathrm{s}}}{\rmd t} = \frac{2\eta_1}{1-2\eta_1}\, U_0, \qquad
  r_{\mathrm{s}} = u_{\mathrm{s}} t.
\end{equation}
Combining equations~(\ref{u_s=}) and (\ref{eta_1=}) and omitting the second and higher order terms with respect to $\eta_1$, we obtain
\begin{equation}\label{eta_1==}
  \eta_1 =\frac{AT_1}{U_0^2+4AT_1}, \qquad \frac{u_{\mathrm{s}}}{U_0} =
  \frac{2AT_1}{U_0^2+2AT_1}.
\end{equation}

The global energy balance for the imploding plasma mass can be expressed as
\begin{equation}\label{glen_bal=}
  P_{\mathrm{X}} +\frac{\rmd}{\rmd t} \left[m_{\mathrm{s}}\epsilon_1 +(m_0-m_{\mathrm{s}}) \frac{U_0^2}{2}
   \right] =0,
\end{equation}
where $P_{\mathrm{X}}$ is the total (per unit cylinder length) power of x-ray emission, which escapes through the outer boundary. If $P_{\mathrm{X}}=0$, we obtain a simple ``conservative'' result
\begin{equation}\label{eps_1,inf=}
  \epsilon_1 =\frac{1}{2}\, U_0^2,
\end{equation}
which yields
\begin{equation}\label{T_1_inf=}
  T_1=T_{1\infty} =\frac{\gamma-1}{2A}\, U_0^2 =
  \left\{ \begin{array}{ll} 1.8 \mbox{ keV} & \mbox{in case A}, \\
  1.3 \mbox{ keV} & \mbox{in case Z}. \end{array} \right.
\end{equation}
When equation~(\ref{eps_1,inf=}) is used with the realistic equation of state of tungsten, provided by the THERMOS code, it yields $T_{1\infty} \approx 0.8$~keV in case~A, and $T_{1\infty} \approx 0.95$~keV in case~Z. It is this post-shock temperature that one would calculate, having literally applied the Raizer's model to a planar stagnation shock in an infinite medium. In our non-conservative situation, where most of the radiation flux escapes the imploding plasma, the final post-shock temperature $T_1$ is significantly lower than $T_{1\infty}$.

To close up our analytical model, we need an expression for $P_{\mathrm{X}}$. If we assume the unshocked infalling plasma to be transparent for the outgoing radiation, we can write
\begin{equation}\label{P_X=}
  P_{\mathrm{X}} = 2\pi r_{\mathrm{s}}\, \sigma T_1^4,
\end{equation}
which means that the opaque compressed core of radius $r_{\mathrm{s}}$  radiates as a black body with a surface temperature $T_1$. Clearly, such a situation should correspond to sufficiently small values of $m_0$, and our case~A, as will be seen below, falls into this category.

An additional approximation that we make when opening the brackets in (\ref{glen_bal=}) is neglect of the term $m_{\mathrm{s}} \rmd\epsilon_1/\rmd t$ compared to $\epsilon_1 \rmd m_{\mathrm{s}}/\rmd t$: this spares us the need to solve a differential equation with practically no loss of accuracy. As a result, upon substitution of (\ref{m_s=}), (\ref{u_s=}), (\ref{eta_1==} and (\ref{P_X=}) into (\ref{glen_bal=}), we arrive at the following equation for determination of $T_1 =T_{1\ast} =T_{1\ast}(t)$
\begin{equation}\label{eq_for_T_1a=}
  T_{1\ast} \left[1+4\pi (\gamma-1) \left(\frac{\Delta_0}{m_0}\right)
  \frac{\sigma T_{1\ast}^4}{U_0^2 +4AT_{1\ast}}\, t\right] = T_{1\infty},
\end{equation}
where $T_{1\infty}$ is given by (\ref{T_1_inf=}). Here we introduced a separate notation $T_{1\ast}$ for the post-shock temperature $T_1$, calculated from (\ref{eq_for_T_1a=}) in the optically thin approximation for the pre-shock plasma, when expression~(\ref{P_X=}) is applied. Having found $T_1=T_{1\ast}$ from (\ref{eq_for_T_1a=}), we calculate $\eta_1$, $u_{\mathrm{s}}$, $r_{\mathrm{s}}$ and $\rho_1$ from equations~(\ref{eta_1==}), (\ref{u_s=}) and (\ref{eta_1-def=}), respectively, and this completes our analytical model for the plasma parameters in the stagnation core.

\begin{figure}[htb!]
\includegraphics*[width=75mm]{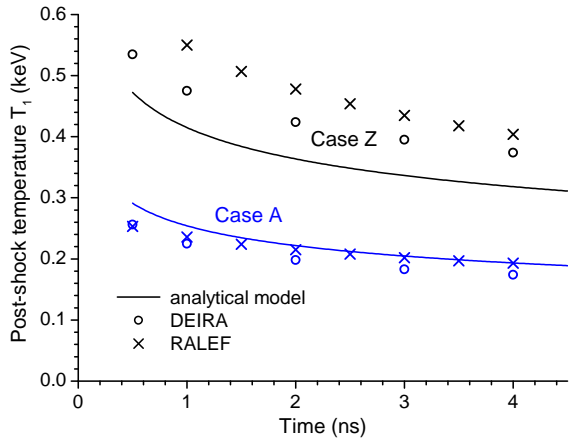}
\caption{\label{f:T1} (Colour online) Time dependence of the post-shock temperature $T_1$: solution of equation~(\ref{eq_for_T_1a=}) $T_1=T_{1\ast} =T_{1\ast}(t)$ (solid curves) is compared with the results of the \mbox{DEIRA} (open circles) and the \mbox{RALEF} (crosses) simulations.}
\end{figure}

Figure \ref{f:T1} compares the values of $T_1=T_{1\ast}$, calculated from equation~(\ref{eq_for_T_1a=}), with those obtained in the \mbox{DEIRA} and \mbox{RALEF} simulations. A very good agreement is observed in case~A, where the optical thickness $\tau_{\mathrm{s}}$ of the pre-shock plasma at different frequencies has moderate values around 1 (see figure~\ref{f:tau-A} below). The agreement becomes worse in case~Z, where $\tau_{\mathrm{s}}$ reaches values around 10 and higher (see figure~\ref{f:tau-Z} below): then equation~(\ref{P_X=}) significantly overestimates the radiative energy loss. From the above analysis it follows that the true value of $T_1$ should be in the range $T_{1\ast} < T_1 < T_{1\infty}$; when $\tau_{\mathrm{s}}>1$ increases, the difference $T_1-T_{1\ast}$ grows and $T_1$ approaches the limiting value of $T_1=T_{1\infty}$. Note that, when considered as a function of the total imploding mass $m_0$ at a fixed value of $U_0$, the post-shock temperature $T_1$ grows with $m_0$ firstly because $T_{1\ast}$ increases [as it follows from (\ref{eq_for_T_1a=})], and, secondly, because the difference $T_1-T_{1\ast}>0$ becomes larger for $\tau_{\mathrm{s}} \gg 1$.

\subsubsection{Temperature peak in the shock front}

\begin{figure}[htb!]
\includegraphics*[width=70mm]{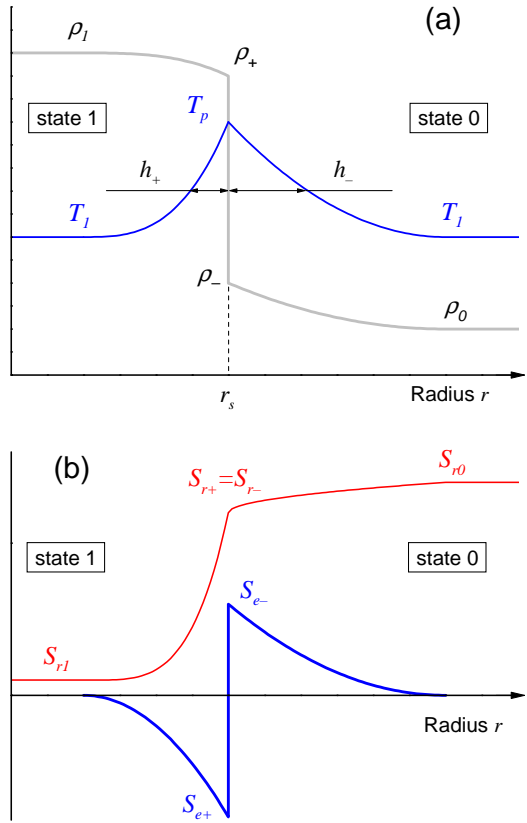}
\caption{\label{f:sh_front} (Colour online) Schematic view of the structure of a supercritical RD shock front with non-zero thermal conduction and zero viscosity. Shown are the density, $\rho$, and the temperature, $T$, profiles (a), as well as the profiles of the conductive, $S_{\mathrm{e}}$, and the radiative, $S_{\mathrm{r}}$, energy fluxes (b). At $r=r_{\mathrm{s}}$ the density $\rho$ and the conductive flux $S_{\mathrm{e}}$ are discontinuous.}
\end{figure}

In addition to the post-shock parameters, one would like to have an estimate for the peak matter (electron) temperature $T_{\mathrm{p}}$ inside the shock front (see figures~\ref{f:deiA} and \ref{f:deiZ}), which defines the hard component of the emitted x-ray spectrum. Such an estimate, however, cannot be obtained without a proper account for thermal conduction. With the conduction energy flux given by \begin{equation}\label{S_e_def=}
  S_{\mathrm{e}} =-\kappa \frac{\partial T}{\partial r},
\end{equation}
where $\kappa$ is the conduction coefficient, the temperature $T$ becomes a continuous function across the density jump, while $S_{\mathrm{e}}$ is discontinuous; the radiation energy flux $S_{\mathrm{r}}$, on the contrary, is everywhere continuous \cite{ZeRa67}. A qualitative view of the density, temperature and the energy flux profiles across a supercritical RD shock front with strong thermal conduction is shown in figure~\ref{f:sh_front}.

In the shock structure shown in figure~\ref{f:sh_front} one can identify four hydrodynamic states: state~0 at the foot of the conduction-preheated layer before the density jump, state~``$-$'' at the entrance into the density jump, state~``$+$'' upon the exit from the density jump, and state~1 behind the post-shock relaxation zone. The effective width (FWHM) of the conduction-preheated layer is $h_-$; the effective width of the post-shock relaxation layer is $h_+$; the effective width of the entire shock front is the sum of the two,
\begin{equation}\label{Dr_s=}
  \Delta r_{\mathrm{s}} =h_- +h_+.
\end{equation}
In our case $h_-$ is determined primarily by thermal conduction, whereas for $h_+$ both the radiant emissivity and thermal conduction are important. Because $h_-$ is much shorter than the shock radius $r_{\mathrm{s}}$, we can assume that the state~0 lies at the end of the broad radiation-preheating zone and, as in the previous subsection, ignore partial plasma compression and deceleration in the latter. Then, the plasma parameters in the four mentioned states can be represented as in table~\ref{t:4states}. The parameters in states 0 and 1 are known from the previous subsection. Here we have to evaluate $h_-$, $h_+$ and $T_{\mathrm{p}}$.

\begin{table}[htb!]
\caption{\label{t:4states} Plasma parameters at four characteristic
states inside the RD shock front.} 
\begin{indented}
\item[]\begin{tabular}{@{}lllll} \br
   & 0     &  ``$-$''   & ``$+$''    & 1     \\ \mr
 $v$     & $v_0 = -(u_{\mathrm{s}}+U_0)$ & $v_- = v_0 \eta_-$ &
           $v_+ = v_1/\eta_+$ & $v_1$ \\
 $\rho$  & $\rho_0 =\rho_0(r_{\mathrm{s}})$  & $\rho_- =\rho_0/\eta_-$ &
           $\rho_+ = \rho_1\eta_+$  & $\rho_1$  \\
 $T$ & $T_1$ & $T_{\mathrm{p}}$ & $T_{\mathrm{p}}$ & $T_1$ \\
$S_{\mathrm{e}}$ & 0 & $S_{\mathrm{e}-}$ & $S_{\mathrm{e}+}$  & 0 \\
 $S_{\mathrm{r}}$  & $S_{\mathrm{r}0}$ & $S_{\mathrm{r}-}= S_{\mathrm{r}+}$ & $S_{\mathrm{r}+}=S_{\mathrm{r}-}$ & $S_{\mathrm{r}1}$ \\ \br
\end{tabular}
\end{indented}
\end{table}

We derive a system of approximate equations for the three unknowns $h_-$, $h_+$ and $T_{\mathrm{p}}$ by successively applying the general equations (\ref{bal-p=}), (\ref{bal-e=}) of momentum and energy balance three times, for transitions between states~0 and ``$-$'', between states~``$-$'' and ``$+$'', and between states ``$+$'' and 1. For each of the three transitions we define a corresponding inverse compression factor
\begin{equation}\label{def_eta_-=}
  \eta_- \equiv \frac{\rho_0}{\rho_-}, \qquad
  \eta_{\mathrm{s}} \equiv \frac{\rho_-}{\rho_+}, \qquad
  \eta_+ \equiv \frac{\rho_+}{\rho_1};
\end{equation}
evidently, we must have
\begin{equation}\label{prod_eta=}
  \eta_-  \eta_{\mathrm{s}} \eta_+ = \eta_1.
\end{equation}

For the first transition between states 0 and ``$-$'' we can neglect the coupling between radiation and matter because here the plasma density $\rho \simeq \rho_0$ is low compared to the compressed state. The latter means that $S_{\mathrm{r}0} \approx S_{\mathrm{r}-}$, and equations~(\ref{bal-p=}), (\ref{bal-e=}) can be written as
\begin{eqnarray}\label{bal-p_1=} 
  A(T_{\mathrm{p}}-T_1\eta_-) = \eta_-(1-\eta_-)\, v_0^2,
  \\ \label{bal-e_1=} 
  \frac{S_{\mathrm{e}-}}{j} = \frac{A\gamma}{\gamma-1} (T_{\mathrm{p}}-T_1) -\frac{v_0^2}{2}
  (1-\eta_-^2).
\end{eqnarray}
Because $T_{\mathrm{p}}-T_1 \lesssim T_1$, we have $1-\eta_- \ll 1$, which enables us to reduce equations~(\ref{bal-p_1=}) and (\ref{bal-e_1=}) to
\begin{eqnarray}\label{eta_-=} 
  1-\eta_- \approx A(T_{\mathrm{p}}-T_1)/v_0^2,
  \\ \label{S_e-=} 
  \frac{S_{\mathrm{e}-}}{j} \approx \frac{A}{\gamma-1} (T_{\mathrm{p}}-T_1).
\end{eqnarray}

The second transition is an isothermal density jump between states ``$-$'' and ``$+$'', where $S_{\mathrm{r}}$ is continuous and $S_{\mathrm{e}}$ jumps from $S_{\mathrm{e}-}$ to $S_{\mathrm{e}+}$. Here equations~(\ref{bal-p=}), (\ref{bal-e=}) take the form
\begin{eqnarray}\label{bal-p2=} 
  AT_{\mathrm{p}}=\eta_{\mathrm{s}} v_-^2 =\eta_{\mathrm{s}} 
  \eta_-^2 v_0^2,
  \\ \label{bal-e2=}  
  \frac{S_{\mathrm{e}+}}{j} =\frac{S_{\mathrm{e}-}}{j} -
   \frac{v_-^2}{2}(1-\eta_{\mathrm{s}}^2).
\end{eqnarray}
Neglecting the second and higher order terms with respect to the small parameters $\eta_{\mathrm{s}}$ and $1-\eta_-$ in (\ref{bal-e2=}), we find \begin{equation}\label{S_e+=}
  \frac{S_{\mathrm{e}+}}{j} \approx \frac{A\gamma}{\gamma-1} (T_{\mathrm{p}}-T_1)
  -\frac{v_0^2}{2}.
\end{equation}

The third transition from state ``$+$'' to state~1 occurs in the compressed state, where the plasma emissivity (roughly proportional to the density $\rho$) is high, and we have to account for variation of the radiation energy flux $S_{\mathrm{r}}$. Hence, equations~(\ref{bal-p=}), (\ref{bal-e=}) take the form
\begin{eqnarray}\label{bal-p3=}
   A(T_1-\eta_+ T_{\mathrm{p}}) = \eta_1^2 v_0^2(\eta_+^{-1}-1),
   \\ \label{bal-e3=}
   \frac{S_{\mathrm{r}+}-S_{\mathrm{r}1}}{j} = \frac{A\gamma}{\gamma-1} (T_{\mathrm{p}}-T_1) +
   \frac{1}{2}\,\eta_1^2v_0^2 \left(\eta_+^{-2} -1\right)
    -\frac{S_{\mathrm{e}+}}{j}.
\end{eqnarray}
Retaining just the leading terms with respect to the small parameter $\eta_1$, we obtain
\begin{eqnarray}\label{eta_+=} 
  \eta_+ \approx T_1/T_{\mathrm{p}} \quad \Leftrightarrow \quad
  \rho_1T_1 \approx \rho_+T_{\mathrm{p}},
  \\ \label{dS_r=}
  S_{\mathrm{r}+} -S_{\mathrm{r}1} \approx \frac{1}{2}\, j v_0^2 
  \approx  \frac{1}{2}\, j U_0^2.
\end{eqnarray}

As a final step, we express the heat conduction fluxes in terms of the corresponding temperature gradients,
\begin{equation}\label{S_e_grad=}
  S_{\mathrm{e}-} \approx \kappa_-\frac{T_{\mathrm{p}}-T_1}{2h_-}, \qquad
  S_{\mathrm{e}+} \approx -\kappa_+\frac{T_{\mathrm{p}}-T_1}{2h_+},
\end{equation}
and the radiation flux increment
\begin{equation}\label{dS_r/k_Pl=}
  S_{\mathrm{r}+}-S_{\mathrm{r}1} \approx \frac{8}{5}\,\sigma k_{\mathrm{Pl}} h_+ \left(T_{\mathrm{p}}^4
  -T_1^4 \right)
\end{equation}
in terms of the post-shock plasma emissivity; in (\ref{S_e_grad=}) $\kappa_-$ and $\kappa_+$ are, respectively, the conduction coefficients in states ``$-$'' and ``$+$''; in (\ref{dS_r/k_Pl=}) $k_{\mathrm{Pl}}$ is the Planckian mean absorption coefficient of radiation in state~``$+$''. Expression~(\ref{dS_r/k_Pl=}) is an approximation to the emission power of an optically thin planar layer, which is valid in both limits of $T_{\mathrm{p}} \gg T_{\mathrm{r}}=T_1$ and $T_{\mathrm{p}} \to T_{\mathrm{r}}=T_1$; the factor $\frac{8}{5}\sigma$ instead of $4\sigma$ takes into account that $h_+$ is the halfwidth of the $T$ rather than the $T^4$ profile. From (\ref{S_e-=}), (\ref{S_e+=}), (\ref{dS_r=})--(\ref{dS_r/k_Pl=}) we obtain the following system of three equations for evaluation of $h_-$, $h_+$ and $T_{\mathrm{p}}$:
\begin{eqnarray}\label{h_-,fin=} 
  h_-=(\gamma-1)\kappa_-/(2jA),
  \\ \label{h_+,fin=} 
  h_+ = \frac{5}{16} \frac{jU_0^2}{\sigma k_{\mathrm{Pl}} (T_{\mathrm{p}}^4-T_1^4)},
  \\ \label{eq_for_T_p=} 
  \frac{jU_0^2}{T_{\mathrm{p}}-T_1}=\frac{\kappa_+}{h_+}+
  \frac{2jA\gamma}{\gamma-1}.
\end{eqnarray}
For numerical estimates we use power-law approximations
\begin{equation}\label{kap_+/-=}
  \kappa_- \approx \kappa_+ \approx 0.15\, T_{\mathrm{p,keV}}^2
  \mbox{ TW cm${}^{-1}$ keV${}^{-1}$},
\end{equation}
\begin{equation}\label{k_Pl=}
  k_{\mathrm{Pl}} \approx 700\, 
  \frac{\rho_{+,\mathrm{g/cc}}}{T_{\mathrm{p,keV}}} \approx 700\, 
\frac{\rho_{1,\mathrm{g/cc}}T_{1,\mathrm{keV}}}{T_{\mathrm{p,keV}}^2}
  \mbox{ cm${}^{-1}$},
\end{equation}
to the THERMOS data for tungsten in the relevant parameter range.

\begin{table}[htb!]
\caption{\label{t:T_p} Comparison of the analytically evaluated stagnation shock parameters with the \mbox{RALEF} results.} \begin{indented} 
\item[]\begin{tabular}{@{}lllll} \br
& \centre{2}{case A, $t=3$~ns} & \centre{2}{case Z, $t=3$~ns} \\ \ns\ns
& \crule{2} &\crule{2} \\
   & analytical & RALEF & analytical & RALEF \\ \mr
 $\rho_1$ (g cm${}^{-3}$)& 4.0 & 3.5 & 14.2 & 7.3 \\
 $T_1$ (keV) &         0.205  & 0.20 & 0.34 & 0.44 \\
 $T_{\mathrm{p}}$ (keV) & 0.406 & 0.35  & 0.65& 0.54 \\
 $\Delta r_{\mathrm{s}}$ ($\mu$m) & 1.2 & 0.8 & 0.3 & $< 0.4$ \\ \mr
 $h_-$ ($\mu$m)& 1.08   & --- & 0.24  & ---  \\
 $h_+$ ($\mu$m)& 0.14  & --- & 0.08  & --- \\ \br
\end{tabular}
\end{indented}
\end{table}

In table~\ref{t:T_p}  the analytically evaluated shock parameters are compared  with those obtained in the \mbox{RALEF} simulations for $t=3$~ns. Generally, the analytical model tends to produce higher values of the peak temperature $T_{\mathrm{p}}$ than the \mbox{DEIRA} and the \mbox{RALEF} codes because of an assumed sharp angle in the temperature profile (see figure~\ref{f:sh_front}), which is smeared either by artificial viscosity in the \mbox{DEIRA} code, or by insufficient spatial resolution in the \mbox{RALEF} simulations.  It is clearly seen that, as one passes from case~A to a more powerful case~Z, the stagnation shock becomes significantly hotter and more narrow --- in full agreement with general properties of supercritical RD shocks \cite{ZeRa67}. Because of the intricate coupling between thermal conduction and radiation emission, no universal power-law scaling for $T_{\mathrm{p}}$ and $\Delta r_{\mathrm{s}}$ can be deduced from (\ref{h_-,fin=})-(\ref{k_Pl=}).

\section{\label{s:ph} X-ray pulse}

The 3T model of the \mbox{DEIRA} code is reasonably adequate for calculating the total power profile of the x-ray pulse (see figures~\ref{f:WrA} and \ref{f:WrZ} below), but can provide no information on its spectral characteristics. For this one has to solve the equation of spectral radiation transfer together with the hydrodynamics equations, and that is where we employ the RALEF-2D code.

\subsection{Power profile}

\begin{figure}[htb!]
\includegraphics*[width=70mm]{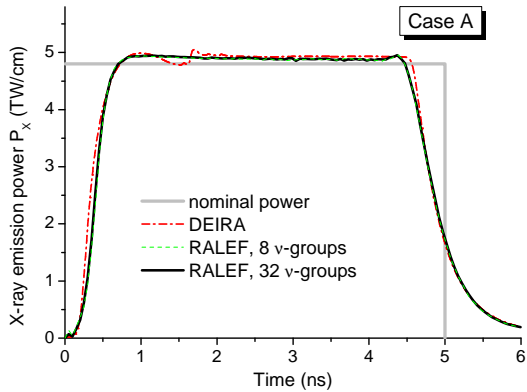}
\caption{\label{f:WrA} (Colour online) Temporal profile of the total x-ray emission power $P_{\mathrm{X}}$ in case~A:  the results of three different numerical simulations are compared among themselves and with the nominal power profile, which corresponds to an instantaneous 100\% conversion of the plasma kinetic energy into x-ray emission.}
\end{figure}

\begin{figure}[htb!]
\includegraphics*[width=70mm]{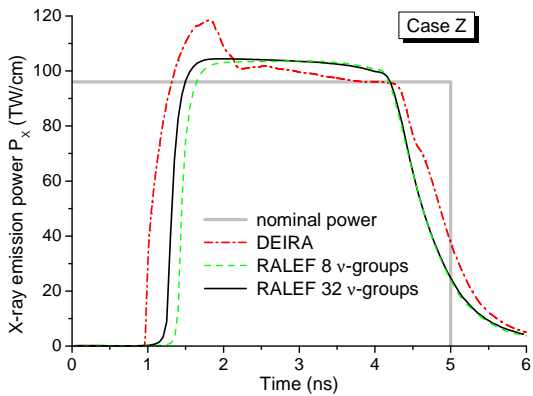}
\caption{\label{f:WrZ} (Colour online) Same as figure~\ref{f:WrA} but in case~Z.}
\end{figure}

Figures \ref{f:WrA} and \ref{f:WrZ} show the temporal x-ray power profiles $P_{\mathrm{X}} =P_{\mathrm{X}}(t)$ as calculated with the \mbox{DEIRA} and the \mbox{RALEF} codes, which agree fairly well with one another, especially in case~A. These profiles demonstrate a clear quasi-steady phase, which lasts about 4~ns in case~A, and about 2.5~ns in case~Z; at this phase $P_{\mathrm{X}}$ is close to the nominal power $P_0$. A marked difference between cases~A and Z is a later (by $\simeq 1$~ns) rise of the x-ray power in case~Z. This delay occurs because in case~Z the radiation heat wave has to propagate through a more massive and optically thick layer of cold plasma before it breaks out to the surface. The overall efficiency of conversion of the initial kinetic energy into radiation (by $t=6$~ns) is 92\% in case~A and 78\% in case~Z according to the \mbox{RALEF} data, and 94\% in case~A and 81\% in case~Z according to the \mbox{DEIRA} results.

\subsection{Shock structure in the RALEF simulations}

\begin{figure}[htb!]
\includegraphics*[width=75mm]{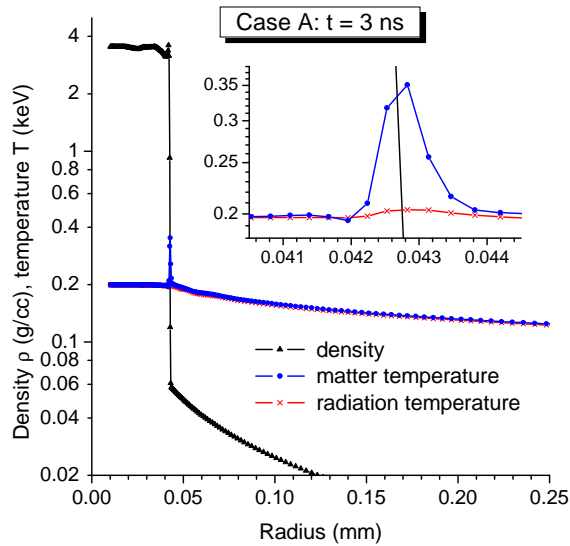}
\caption{\label{f:RAL_Trho_A} (Colour online) Radial density and temperature profiles at $t=3$~ns in case~A obtained in the 2D \mbox{RALEF} simulation with 32 spectral groups.}
\end{figure}

\begin{figure}[htb!]
\includegraphics*[width=75mm]{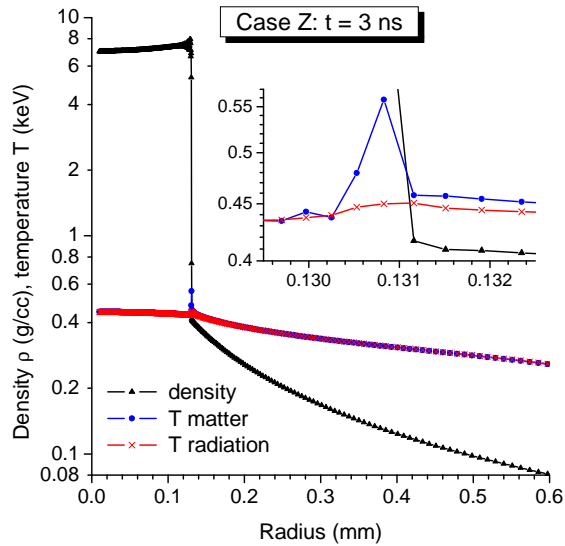}
\caption{\label{f:RAL_Trho_Z} (Colour online) Same as figure~\ref{f:RAL_Trho_A} but in case~Z.}
\end{figure}

Radial density and temperature profiles in the imploding plasma, obtained with the \mbox{RALEF} code, are shown in figures~\ref{f:RAL_Trho_A} and \ref{f:RAL_Trho_Z}. Despite quite different physical models, the \mbox{RALEF} and the \mbox{DEIRA} results agree almost perfectly in case~A: we calculate practically the same values of the post-shock, $T_1$, and the peak, $T_{\mathrm{p}}$, matter temperatures. Figure~\ref{f:RAL_Trho_A} also demonstrates that in case~A the temperature peak is fairly well resolved in 2D simulations, although appears somewhat broader than in the 1D \mbox{DEIRA} picture. Larger 2D values of the shock radius $r_{\mathrm{s}}$ are explained by different position of the inner boundary (at $r=10$~${}\mu$m in the 2D case versus $r=0$ in the 1D case).

In case~Z, on the contrary, the temperature peak is rather poorly resolved in 2D simulations, as one sees in figure~\ref{f:RAL_Trho_Z} --- despite a larger total number of radial mesh zones (600 in case~Z versus 250 in case~A). The reason is twofold: on the one hand, the temperature peak in case~Z is about a factor 2 more narrow than in case~A; on the other, a considerably larger shock radius $r_{\mathrm{s}}$ causes the RALEF mesh rezoning algorithm to force a coarser grid along the radial direction. Nevertheless, the agreement between the \mbox{RALEF} and the \mbox{DEIRA} results for the post-shock, $T_1$, and the peak, $T_{\mathrm{p}}$, matter temperatures is also fairly good.

\begin{figure}[htb!]
\includegraphics*[width=75mm]{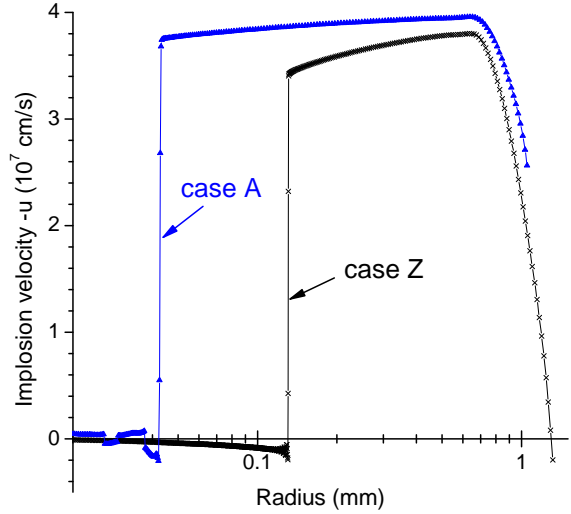}
\caption{\label{f:u} (Colour online)  Radial profiles of the plasma implosion velocity (minus the radial velocity $u$) at $t=3$~ns in cases~A and Z as calculated by the \mbox{RALEF} code with 32 spectral groups. Logarithmic scale for the radius allows to show the detailed structure of the compact shocked region together with the overall large scale behaviour.}
\end{figure}

Radial profiles of the implosion velocity $-u(r)$ at $t=3$~ns are displayed in figure~\ref{f:u} for both cases A and Z. One notices that the fluid velocity changes sign across the shock front. As was already mentioned in section~\ref{s:score}, the post-shock plasma on average slowly expands (i.e.\ has a negative implosion velocity) because the stagnation shock propagates over a falling density profile. Near the outer edge, the infalling plasma has already been significantly decelerated, especially in the more massive case~Z. The decelerating pressure gradient is created by re-deposition of radiant energy transported from the stagnation shock front.

\begin{figure}[htb!]
\includegraphics*[width=75mm]{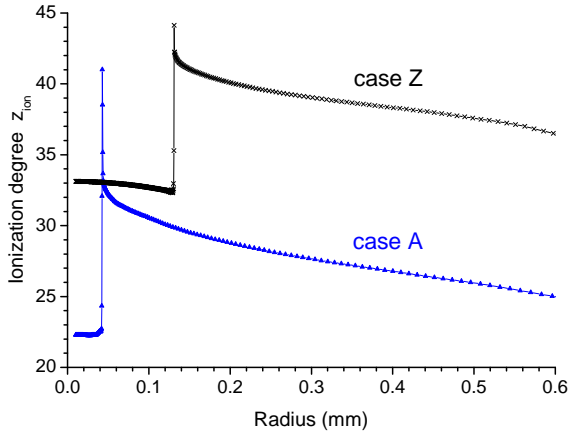}
\caption{\label{f:zion} (Colour online) Radial profiles of the tungsten ionization degree $z_{\mathrm{ion}}$ at $t=3$~ns in cases~A and Z  as calculated by the \mbox{RALEF} code with 32 spectral groups in the radiation transport module.}
\end{figure}

Figure \ref{f:zion} shows spatial profiles of the mean ion charge $z_{\mathrm{ion}}$ at $t=3$~ns. One sees that tungsten ions with charges of $z_{\mathrm{ion}} = 40$--45 are present inside the stagnation shock front.  It should be reminded here that these ion charges have been calculated in the LTE limit. A direct evidence that the LTE approximation is quite adequate in our situation is a close agreement between the radiation and matter temperatures in figures~\ref{f:RAL_Trho_A} and \ref{f:RAL_Trho_Z}. The applicability of LTE can only be questioned inside the narrow shock front. However, the plasma density there is already so high ($n_{\mathrm{e}} \gtrsim 6\times 10^{21}$ in case~A, and $n_{\mathrm{e}} \gtrsim 5\times 10^{22}$ in case~Z) that non-LTE corrections to the values of $z_{\mathrm{ion}}$ and $T$ inside the shock front are not expected to be significant (poor spatial resolution of this region may, in fact, be a no less important issue).

\subsection{Spatially integrated spectra}

\subsubsection{Case A}

\begin{figure}[htb!]
\includegraphics*[width=75mm]{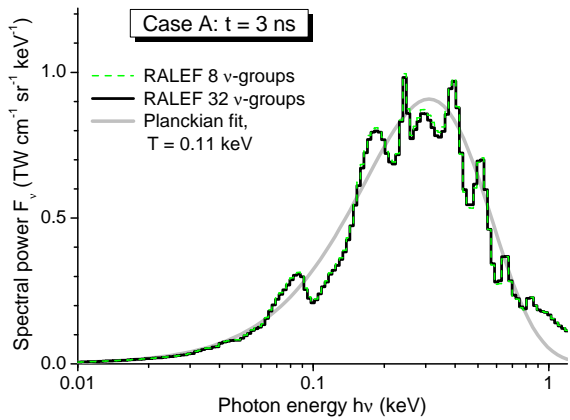}
\caption{\label{f:totsp_A-lin} (Colour online)  Spectral power of x-ray emission per unit cylinder length at $t=3$~ns in case~A: the soft part of the x-ray spectrum. The Planckian-fit curve is normalized to the emission peak.} 
\end{figure}

\begin{figure}[htb!]
\includegraphics*[width=75mm]{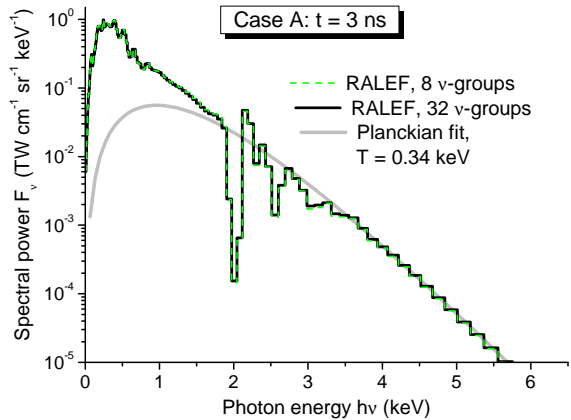}
\caption{\label{f:totsp_A-exp} (Colour online) Spectral power of x-ray emission per unit cylinder length $t=3$~ns in case~A: the hard part of the x-ray spectrum. The Planckian-fit curve is normalized to the $h\nu \gtrsim 3$~keV tail of the emission.}
\end{figure}

The overall x-ray spectrum emitted by the imploding pinch in case~A at $t=3$~ns is shown in two different representations in figures~\ref{f:totsp_A-lin} and \ref{f:totsp_A-exp}. This spectrum would have been observed through an imaginary slit perpendicular to the pinch axis by a detector without spatial resolution. More precisely, figures~\ref{f:totsp_A-lin} and \ref{f:totsp_A-exp} display the spectral power $F_{\nu}$ [TW cm${}^{-1}$ sr${}^{-1}$ keV${}^{-1}$] per unit cylinder length, obtained by integrating along the slit the intensity $I_{\nu}(\mathbf{\Omega})$ of the outgoing radiation, which propagates in direction $\mathbf{\Omega}$ perpendicular to the pinch axis. The shown spectrum was obtained by solving the transfer equation (\ref{eq:r-transp=}) in the post-processor mode for 200 spectral groups of the secondary frequency set. In case~A it turns out to be rather insensitive to the number of spectral groups (either 8 or 32) coupled to the hydrodynamics energy equation.

The plots in figures~\ref{f:totsp_A-lin} and \ref{f:totsp_A-exp} demonstrate that the emitted spectrum can roughly be approximated as a superposition of two Planckian curves: one with a temperature $T_{\mathrm{r}1} \approx 0.11$~keV, and the other with a temperature $T_{\mathrm{r}2} \approx 0.34$~keV. The interpretation of the hard component is straightforward: it is the thermal emission of the temperature peak $T_{\mathrm{p}}=T_{\mathrm{ep}} \approx T_{\mathrm{r}2}$ inside the stagnation shock. In our case this component carries about 16\% of the total x-ray flux and is emitted by an optically thin plasma layer rather than by a surface of a black body.

The soft component originates from a much broader halo around the shock front, at an effective radius of $r_{\mathrm{em}} \approx 0.4$~mm${} \gg r_{\mathrm{s}} = 0.043$~mm. This halo is the result of reprocession  of the original shock emission by the cold layers of the unshocked material. Note that the temperature $T_{\mathrm{r}1}$ of the soft component is significantly lower than the post-shock matter temperature $T_1 = 0.20$~keV, which implies that even in the low-mass case~A the infalling unshocked plasma is not truly optically thin.

\begin{figure}[htb!]
\includegraphics*[width=75mm]{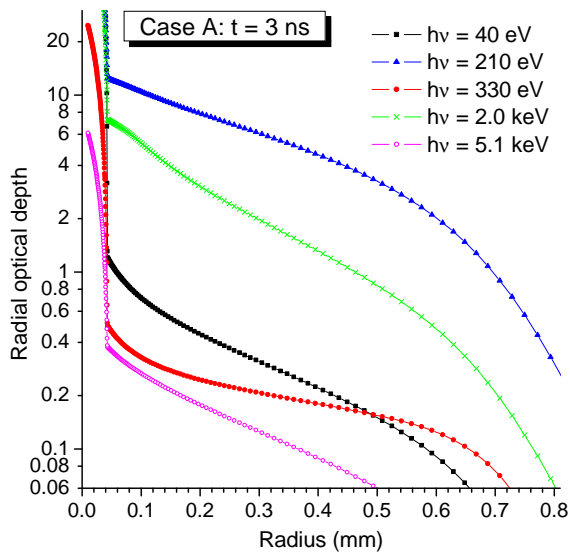}
\caption{\label{f:tau-A} (Colour online) Profiles of radial optical thickness at different photon energies for $t=3$~ns in case~A.}
\end{figure}

Figure \ref{f:tau-A} provides more detailed information on the radial profiles of the spectral optical depth. It is seen that, depending on the photon energy, the optical thickness of the unshocked plasma can be either significantly below or significantly above unity. The latter means that the effective emitting layer is, in fact, not well defined, and the observed spectrum may exhibit significant deviations from the Planckian shape. Indeed, a number of prominent dips and spikes in the calculated spectrum in figures~\ref{f:totsp_A-lin} and \ref{f:totsp_A-exp} arise as a combined effect of a complex spectral dependence of the tungsten opacity, shown in figure~\ref{f:knu}, superimposed on a nontrivial temperature distribution inside and above the stagnation shock.

\subsubsection{Case Z}

\begin{figure}[htb!]
\includegraphics*[width=75mm]{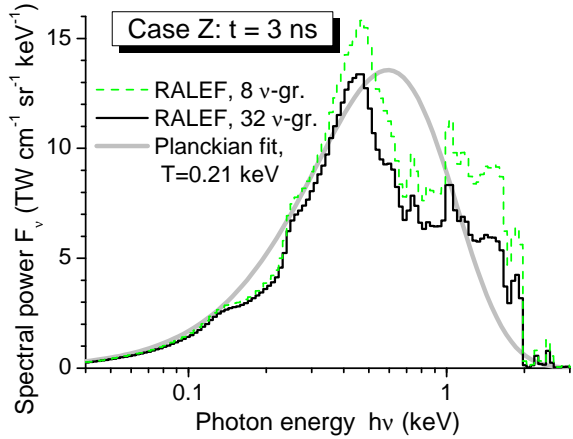}
\caption{\label{f:totsp_Z-lin} (Colour online) Same as figure~\ref{f:totsp_A-lin} but for case~Z.}
\end{figure}

\begin{figure}[htb!]
\includegraphics*[width=75mm]{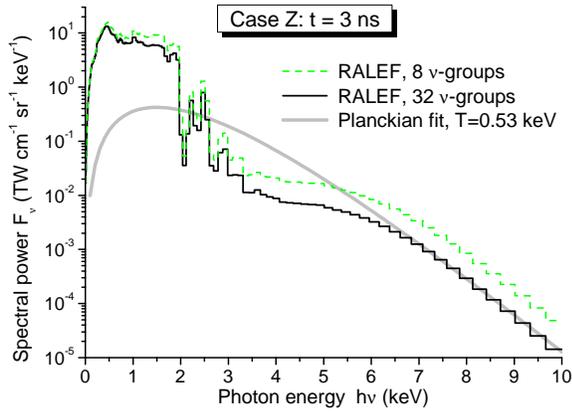}
\caption{\label{f:totsp_Z-exp} (Colour online) Same as figure~\ref{f:totsp_A-exp} but for case~Z.  The Planckian-fit curve is normalized to the $h\nu \gtrsim 6$~keV tail of the emission.}
\end{figure}

Figures \ref{f:totsp_Z-lin} and \ref{f:totsp_Z-exp} display the same information as figures~\ref{f:totsp_A-lin} and \ref{f:totsp_A-exp} but for a 20 times larger (6~mg~cm${}^{-1}$) imploding mass of case~Z. Here both the main component of the spectrum in figure~\ref{f:totsp_Z-lin} and the hard component in figure~\ref{f:totsp_Z-exp} correspond to roughly two times higher Planckian-fit temperatures of $T_{\mathrm{r}1} =0.21$~keV and $T_{\mathrm{r}2} =0.53$~keV; the hard component carries about 7\% of the total x-ray flux.

\begin{figure}[htb!]
\includegraphics*[width=75mm]{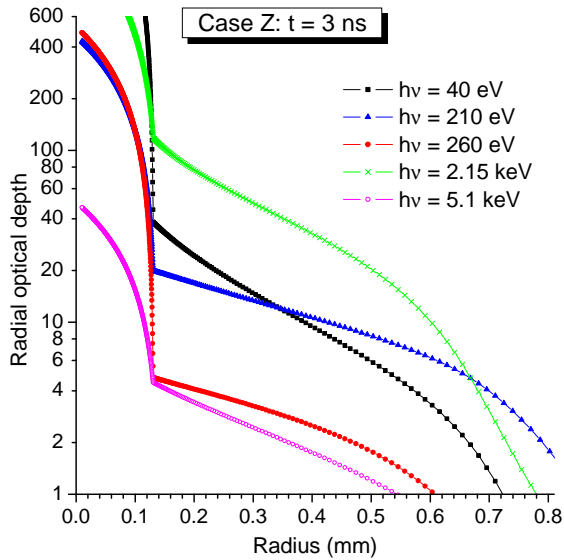}
\caption{\label{f:tau-Z} (Colour online) Same as figure~\ref{f:tau-A} but for case~Z.}
\end{figure}

In contrast to case~A, now the shock front lies at an optical depth $\tau_{\mathrm{s}}$ well in excess of unity at all frequencies: as can be seen in figure~\ref{f:tau-Z}, at $t=3$~ns the optical depth $\tau_{\mathrm{s}}$ varies in the range $\tau_{\mathrm{s}} \approx 4$--100. As a result, the calculated spectrum in figures~\ref{f:totsp_Z-lin} and \ref{f:totsp_Z-exp} demonstrates higher sensitivity to the number of spectral groups coupled to hydrodynamics. The effective emission radius for the equivalent Planckian flux can be evaluated as $r_{\mathrm{em}} \approx 0.7$~mm. Figure~\ref{f:tau-Z} shows that it is around this radius that the spectral optical depth is on the order of unity.

Our calculated spectrum in figure~\ref{f:totsp_Z-exp} appears to be in a fair agreement with the observed x-ray spectra for 6~mg~cm${}^{-1}$ tungsten arrays tested on the Z machine \cite{CuWa.05,CuVe.06}, although the published experimental data at $h\nu \gtrsim 3$--4~keV are rather scarce.  In fact, when we superpose our spectrum in figure~\ref{f:totsp_Z-exp} on that from \cite{FoHe.04}, we observe a very good agreement without even rescaling the absolute fluxes.  The experimental points for $h\nu > 2$~keV, quoted in \cite{CuVe.06,FoHe.04}, do indicate the presence of a hard x-ray component with an effective temperature of $T_{\mathrm{r}2} \approx 0.6$~keV, whereas the main emission is reasonably well approximated by a blackbody spectrum with $T_{\mathrm{r}1} \approx 165$~eV \cite{FoHe.04}. Note that, according to our results, particularly in the region $h\nu = 3$--6~keV, the spectral slope  appears to be significantly flattened as compared to the corresponding Planckian fit of the hard component --- which implies complications for any direct interpretation of the Planckian-fit temperature, inferred from the experimental data in this region.

\subsection{Calculated 1D x-ray images}

\begin{figure}[htb!]
\includegraphics*[width=75mm]{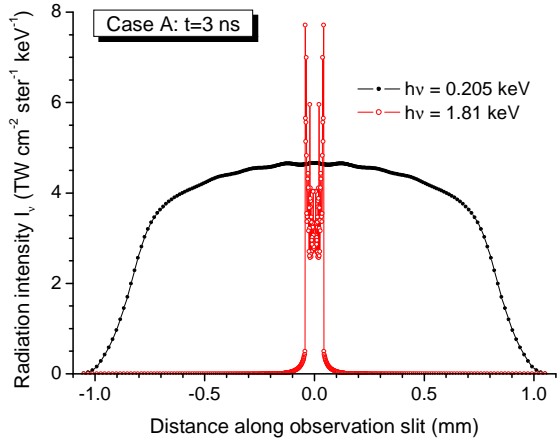}
\caption{\label{f:imgA} (Colour online) 1D x-ray image of the imploding pinch at two different frequencies in case~A at $t=3$~ns.}
\end{figure}

\begin{figure}[htb!]
\includegraphics*[width=75mm]{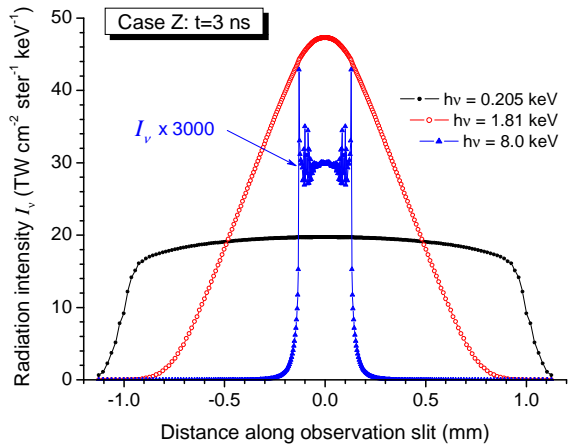}
\caption{\label{f:imgZ} (Colour online) Same as figure~\ref{f:imgA} but for case~Z and three different frequencies.}
\end{figure}

Beside spatially integrated emission spectra, of certain interest might be theoretical spectral images of the imploding pinch.  A selection of such images is shown in figures~\ref{f:imgA} and \ref{f:imgZ} for the time $t=3$~ns. Here the radiation intensity $I_{\nu}= I_{\nu}(s,\mathbf{\Omega})$ is plotted as a function of distance along an imaginary observation slit, perpendicular to the pinch axis, as it would be registered by an observer at infinity; the photon propagation direction $\mathbf{\Omega}$ is also perpendicular the pinch axis. Again, these images have been constructed in the post-processor mode by separate integration of the transfer equation (\ref{eq:r-transp=}) along a predefined set of rays (long characteristics) at selected photon energies. This enabled us to get rid of the numerical diffusion inherent in the method of short characteristics.

Our principal motivation for presenting the images in figures~\ref{f:imgA} and \ref{f:imgZ} was to illustrate how one could possibly resolve the RD shock front, buried deeply inside the imploding plasma column. Figure \ref{f:imgA} demonstrates that in the low-mass case~A this could already be achieved by radiography at photon energies around $h\nu \approx 2$~keV. In the more massive case~Z one has to do the measurements in harder x-rays at $h\nu \gtrsim 8$~keV. The softer part of the spectrum reveals only a broad blurred halo from the imploding plasma, whose size depends on the observation frequency.

\section{Summary}

In this work, within the framework of pure radiation hydrodynamics, we have attempted to present a detailed physical picture of how the kinetic energy of the imploding high-Z (tungsten) plasma in wire array z-pinches could be converted into powerful bursts of x-rays. Having concentrated on a self-consistent modelling of the emergent x-ray spectra, we adopted the simplest possible formulation of the problem. In particular, we assumed that at the final stage of kinetic energy dissipation the dynamic effects due to the magnetic field could be neglected, and that the imploding tungsten plasma has a perfectly symmetric one-dimensional cylindrical configuration. Both assumptions imply severe idealization of the problem, and how realistic are the conclusions reached under them, remains to be clarified by future work.

The reason for our 1D statement of the problem is simply because the 1D picture is always a necessary starting point when exploring a complex physical phenomenon: later on it may serve as a valuable reference case --- especially if it manages to capture the basic physical features of the studied phenomenon. 

However, even if we skip the initial phase of plasma acceleration by the  $\mathbf{j} \times \mathbf{B}$ force and stay within the 1D picture, there remains a question of possible dynamic and kinetic effects due to the (partially) frozen-in magnetic field. We do not expect that such effects can significantly alter the present physical picture of the x-ray pulse formation (at least not in the phase of what we call the main x-ray pulse) simply because the initial Alfvenic Mach number is very high ($\gtrsim 40$). Later on, as the bulk of the imploding mass passes through the stagnation shock and the pinch enters the stagnation phase with a Bennet-type equilibrium, the effects due to the magnetic field and the ensuing MHD instabilities may, of course, become much more significant. This second phase of the x-ray pulse, which may in fact account for a large portion of the total emitted x-ray energy and be strongly dominated by the MHD effects, was not the topic of our present work.

Within the approximations made, we have demonstrated that the conversion of the implosion energy into quasi-thermal x-rays occurs in a very narrow (sub-micron) radiation-dominated shock front, namely, in an RD stagnation shock with a supercritical amplitude according to the classification of \cite{ZeRa67}. We analyzed the structure of the stagnation RD shock by using two independent radiation-hydrodynamics codes, and by constructing an approximate analytical model.

We have found that the x-ray spectrum, calculated with the 2D RALEF code by solving the equation of spectral radiative transfer in the imploding plasma, agrees fairly well  with the published experimental data for the 6~mg~cm${}^{-1}$ tungsten wire arrays tested at Sandia. The hard component of the x-ray spectrum with a blackbody temperature of $T_{\mathrm{r}2} \approx 0.5$--0.6~keV is shown to originate from a narrow peak of the electron temperature inside the RD stagnation shock. Our approximate model clarifies how the width and the amplitude of this temperature peak depend on the imploding plasma parameters. The softer main component of the x-ray pulse is generated in an extended halo around the stagnation shock, where the primary emission from the shock front is absorbed and reemitted by the outer layers of the imploding plasma.

In reality, due to flow non-uniformities, the narrow front of the stagnation shock will almost certainly have a much more irregular shape than in the present 1D picture. But its main characteristics --- the transverse thickness $\Delta r_{\mathrm{s}}$ and the peak electron temperature $T_{\mathrm{ep}}$ --- are controlled by the plasma flow parameters (the implosion velocity $U_0$, the mass flux density $\rho_0 U_0$, the plasma thermal conductivity $\kappa$ and the spectral absorption coefficient $k_{\nu}$) that are not expected to be dramatically affected by moderate flow perturbations. Hence, we expect that radiation-hydrodynamics simulations of realistically perturbed implosions should produce emergent x-ray spectra close to those calculated in the present work.

\ack

This work was partially supported by the Russian Foundation for Basic Research (grant No.~09-02-01532-a), and  by the ExtreMe Matter Institute EMMI in the framework of the Helmholtz Alliance Program HA216/EMMI.

\end{document}